\documentclass[onecolumn,showpacs,nofootinbib,11pt]{revtex4-1}
\usepackage{color}
\usepackage{amssymb,amsmath,yfonts,upgreek}

\newcommand{\ba}{\begin{eqnarray}}
\newcommand{\ea}{\end{eqnarray}}
\newcommand{\bege}{\begin{equation}}
\newcommand{\enge}{\end{equation}}

\usepackage[colorlinks,hyperindex]{hyperref}
\newcommand{\beq}{\begin{eqnarray}}
\newcommand{\benu}{\begin{enumerate}}
\newcommand{\enu}{\end{enumerate}}
\newcommand{\eeq}{\end{eqnarray}}

\newcommand{\RR}{\mathbb{R}}
\newcommand{\mt}{\mathcal}
\newcommand{\cl}{\mt{C}\ell}

\newcommand{\CC}{\mathbb{C}}

\usepackage{hyperref}

\begin{document}

\title{On the Spinor Representation}

\author{J. M. Hoff da Silva}
\email{hoff@feg.unesp.br} \affiliation{Departamento de F\'{\i}sica e Qu\'{\i}mica, Universidade Estadual Paulista, Guaratinguet\'{a}, SP, Brazil}

\author{Rold\~ao da Rocha}
\email{roldao.rocha@ufabc.edu.br}
\affiliation{Centro de Matem\'atica, Computa\c c\~ao e Cogni\c c\~ao, Universidade Federal do ABC - UFABC\\ 09210-580, Santo Andr\'e, Brazil.}
\author{C. H. Coronado Villalobos}
\email{ccoronado@feg.unesp.br} \affiliation{Departamento de F\'{\i}sica e Qu\'{\i}mica, Universidade Estadual Paulista, Guaratinguet\'{a}, SP, Brazil}

\author{R. J. Bueno Rogerio}
\email{rodolforogerio@feg.unesp.br} \affiliation{Departamento de F\'{\i}sica e Qu\'{\i}mica, Universidade Estadual Paulista, Guaratinguet\'{a}, SP, Brazil}


\begin{abstract}
A systematic study of the spinor representation by means of the fermionic physical space is accomplished and implemented. The spinor representation space is shown to be constrained by the Fierz-Pauli-Kofink identities among the spinor bilinear covariants. A robust geometric and topological structure can be manifested from the spinor space, wherein the first and second homotopy groups play prominent roles on the underlying physical properties, associated to fermionic fields.
\end{abstract}
\maketitle

\section{Introduction}

The very definition of a spinor in dealing with physics may be treated as a matter of some importance itself whatsoever. In fact, from simple quaternionic compositions revealing a definite rotation \cite{ALT} to the fermionic quantum internal structure \cite{WEI}, the spinorial approach reveals its richness. Among these possible systematizations concerning spinors, there is a particularly relevant one that encodes all the algebraic necessary information and the important relativistic construction as well, namely, the multivector spinor representation. When represented as a section of a bundle comprised by the $SL(2,\mathbb{C})$ group and $\mathbb{C}^4$, it is possible to understand several spinor properties by inspecting the multivector part constructed out specific $SL(2,\mathbb{C})$ objects. These objects are nothing but the bilinear covariants associated to the regarded spinor  \cite{LOU,Cra}.

Following this reasoning, it is not surprising the usefulness of such a representation, since the bilinear covariants are, at least in principle, related to a set of fermionic observables. Our aim in this paper is to delineate the importance of the representation space, by studying its properties, and then relating them to their physical  consequences. As one will realize, the representation space is quite complicated due to the constraints coming out the Fierz-Pauli-Kofink identities. However, a systematic study of the space properties ends up being useful to relate different domains (subspaces) to the corresponding physics. Moreover, this study allows to face the fermions from a different and useful perspective. 

This paper is organized as follows: in the next section the standard framework and the three equivalent definitions of spinors are revisited for the Minkowski spacetime, emphasizing the most relevant aspects concerning our purposes. Sect. III is devoted to approach the Lounesto's spinors classification and related issues. In Sect. IV we construct and study the spinor representation space and explore the topological and physical consequences. In the final section we conclude.

\section{The three equivalent definitions of spinors}

Consider the Minkowski spacetime $(M\simeq\RR^4, \eta_{\mu\nu})$ and its tangent bundle $
TM$, where $\eta$ denotes the Minkowski metric and Greek (spacetime) indexes run from 0 to 3. Denoting sections of the exterior bundle by
$\sec\Omega (M)$,  the spacetime Clifford algebra shall be denoted by $\cl_{1,3}$.  The set $\{{e}_{\mu }\}$ represents sections of the frame bundle
$\mathbf{P}_{\mathrm{SO}_{1,3}^{e}}(M)$, whereas the set  $\{\gamma^{\mu }\}$ can be further thought as being  the dual
basis, $\gamma^{\mu }({e}_{\nu})=\delta^\mu_{\;\nu}$.
 Classical spinors are objects of the space that carries the usual
$\tau=(1/2,0)\oplus (0,1/2)$ representation of the Lorentz group, that  can be thought as being sections of the vector bundle
$\mathbf{P}_{\mathrm{Spin}_{1,3}^{e}}(M)\times _{\tau }\mathbb{C}^{4}$ \cite{Mosna:2002fr,50}.

 The underlying idea that can join the three definitions of spinors relies on a quite straightforward root, and was inspired by the spacetime algebra, whose elements satisfy $e_\mu e_\nu + e_\nu e_\mu = 2\eta_{\mu\nu}\mathbf{1}$. Indeed,  any arbitrary element 
$\Pi = s+s^{\mu}{e}_{\mu}+s^{\mu\nu}{e}_{\mu}e_{\nu}+s^{\mu\nu\tau}{e}_{\mu}e_{\nu}e_{\tau}+p{e}_{0}e_{1}e_{2}e_{3}\in {\mathcal{C}}\ell _{1,3}
$ 
has a quaternionic representation.  By denoting $\mathbb{H}$ the quaternionic ring, a spinor representation of the Clifford algebra ${\mathcal{C}}\ell _{1,3}\simeq {\mathcal{M}}(2,\mathbb{H})$ can be derived. A primitive idempotent
$f=\frac{1}{2}(1+{e}_{0})$  defines a minimal left ideal ${\mathcal{C}}\ell _{1,3}f$, whose arbitrary element can be expressed as \cite{jayme,jayme1}\begin{eqnarray}
 \upxi &=&\left[\left(s+s^{0}\right)+\left(s^{23}+s^{023}\right){e}_{2}e_{3}-\left(s^{13}+s^{013}\right){e}_{3}e_{1}+\left(s^{12}+s^{012}\right){e}_{1}e_{2}\right]f\nonumber\\&&+\left[\left(
p -s^{123}\right)+\left(s^{1}-s^{01}\right){e}_{2}e_{3}+\left(s^{2}-s^{02}\right){e}_{3}e_{1}+\left(s^{3}-s^{03}\right)\right]{e}_{0}e_{1}e_{2}e_{3}f,
\label{kkk}
\end{eqnarray}
\noindent
constituting then an algebraic spinor $\upxi \in \cl_{1,3}f$. 
The set comprised by the units $\textgoth{i}={e}_{2}{e}_{3},\;\textgoth{j}={e}_{3}{e}_{1},$ and $\textgoth{k}={e}_{1}{e}_{2}$ settles a basis for the quaternionic algebra~$\mathbb{H}$.

Representations of the $\{{e}_{\mu}\}$ in $\mathcal{M}(2,\mathbb{H})$ read 
 \cite{abla}: 
\begin{equation}\label{quatrn}
\left[{e}_{1}\right]=%
{\scriptstyle\begin{pmatrix}
0 & \textgoth{i} \\
\textgoth{i} & 0%
\end{pmatrix}}%
,\quad\left[{e}_{2}\right]=%
{\scriptstyle\begin{pmatrix}
0 & \textgoth{j} \\
\textgoth{j} & 0%
\end{pmatrix}}%
,\quad\left[{e}_{3}\right]=%
{\scriptstyle\begin{pmatrix}
0 & \textgoth{k} \\
\textgoth{k} & 0%
\end{pmatrix}},\quad \left[{e}_{0}\right]=%
{\scriptstyle\begin{pmatrix}
1 & 0 \\
0 & -1%
\end{pmatrix}}\,.%
\end{equation}%
\noindent
Then, the elements $f$ and ${e}_{0}e_{1}e_{2}e_{3}f$ have, respectively, the  representations 
$
[f]={\footnotesize\left(\begin{matrix}
1 & 0 \\
0 & 0%
\end{matrix}\right)}$ and $[{e}_{0}e_{1}e_{2}e_{3}f]=%
{\footnotesize\begin{pmatrix}
0 & 0 \\
1 & 0%
\end{pmatrix}} 
$   \cite{oxford}. 
Hence, an arbitrary element $\Psi \in \cl_{1,3}^{+}$ in the even subalgebra, corresponds  to the so called {spinor operator} \cite{LOU} 
\begin{equation}
\Psi=s+s^{\mu\nu}{e}_{\mu}e_{\nu}+p{e}_{0}e_{1}e_{2}e_{3} 
\label{400}\simeq 
{\scriptstyle\begin{pmatrix}
q_{1} & -q_{2} \\
q_{2} & q_{1}%
\end{pmatrix}}\in\mathcal{M}(2,\mathbb{H}),
\end{equation}
where, according to Eq. (\ref{kkk}), it yields
\begin{eqnarray}
q_{1}&=&s+s^{23}\textgoth{i}+s^{31}\textgoth{j}+s^{12}\textgoth{k}\,,\\
q_2&=&-p+s^{01}\textgoth{i}+s^{02}\textgoth{j}+s^{03}\textgoth{k}\,.
\end{eqnarray}
The vector space isomorphisms $\mathbb{H}^{2}\simeq{\mathcal{C}}\ell _{1,3}^{+} \simeq \mathbb{C}^{4}\simeq {\mathcal{C}}\ell _{1,3}f$ constitute the landmark for the relationship among the spinor operator, the algebraic, and the classical definitions of a spinor \cite{50,Figueiredo:1990bw}.  Hence, it is possible to alternatively write the Dirac algebraic spinor field as an element ${\scriptstyle\begin{pmatrix} q_1\\q_2 \end{pmatrix}}$ of the ring $\mathbb{H}\oplus \mathbb{H}$, as \cite{jayme,jayme1,abla}
\begin{equation}\begin{pmatrix}
q_{1} & -q_{2} \\
q_{2} & q_{1}%
\end{pmatrix} [f]\in \cl_{1,3}f.  \label{hh}
\end{equation}%
\noindent
Returning to Eq.~(\ref{400}), and using for instance the standard representation, 
the complex matrix associated to the spinor operator $\Psi$ in~(\ref{400})  reads \begin{equation}
[\Psi]=\begin{pmatrix}
\mathbb{A}&\mathbb{-B}\\
\mathbb{B}&\mathbb{A}\end{pmatrix},\quad\qquad \text{for} \quad 
\mathbb{A}=
\begin{pmatrix}
\psi_1 & -\psi_2^{*} \\
\psi_2 & \phantom{-}\psi_1^{*} 
\end{pmatrix},\quad\mathbb{B}=
\begin{pmatrix}
\psi_3 & -\psi_4^{*}\\
\psi_4 & \phantom{-}\psi_3^{*}
\end{pmatrix},
\end{equation}
where 
\begin{eqnarray}
\psi_1=s+s^{23}i,\qquad
\psi_2=s^{13}+s^{12}i\qquad
\psi_3=p+s^{10}i,\qquad
\psi_4=s^{02}+s^{30}i
\end{eqnarray}
The standard Dirac spinor $\psi$ was identified, e. g., in Ref. \cite{LOU} as an element of the minimal left ideal $(\mathbb{C}\otimes \cl _{1,3}){\rm f}$ associated to the complexified spacetime algebra $(\mathbb{C}\otimes \cl _{1,3})$, generated by the primitive idempotent \cite{LOU} 
\begin{eqnarray}
{\rm f}=\frac{1}{4}(1+e_{0})(1+ie_{1}e_{2})=\begin{pmatrix}
1 & 0 & 0 & 0 \\
0 & 0 & 0 & 0 \\
0 & 0 & 0 & 0 \\
0 & 0 & 0 & 0
\end{pmatrix},\end{eqnarray} yielding $\psi =\Pi \frac{1}{2}(1+i\gamma _{1}\gamma_{2})\in (\mathbb{C}\otimes \cl _{1,3})f$, with the identification $e_\mu\mapsto \gamma_\mu$. It yields the  bijection between the algebraic spinor \cite{Mosna:2002fr,50,hestenes}
\begin{equation}
\psi =
\begin{pmatrix}
\psi_1 & 0 & 0 & 0 \\
\psi_2 & 0 & 0 & 0 \\
\psi_3 & 0 & 0 & 0 \\
\psi_4 & 0 & 0 & 0
\end{pmatrix}\in (\CC\otimes \cl _{1,3})f \simeq \mathcal{M}(4,\mathbb{C}), \end{equation} and the classical one $\psi=(\psi_1, \psi_2, \psi_3, \psi_4)^\intercal
\in \mathbb{C}^{4}.$ 

Given a representation $\rho: \mathbb{C}\otimes \cl _{1,3}\to \mathcal{M}(4,\CC)$, the adjoint of $A\in\mathbb{C}\otimes \cl _{1,3}$, defined by $A^\dagger = \rho^{-1}(\rho(A)^\dagger)$ (where $\rho(A)^\dagger$ denotes the standard Hermitian conjugation in $\mathcal{M}(4,\CC)$),
reads $A^\dagger = e_0\tilde{A}^* e_0$, where $\tilde{A}$ stands for the reversion of $A$ and $(\,\cdot\,)^*$ denotes the complex conjugation. Besides, its trace is given by Tr$(\rho(\psi)) = 4\langle \psi\rangle_0$, where this notation is used to indicate the projection of a multivector onto 
its scalar part. 

This correspondence provides an immediate identification between $\psi$ and the classical Dirac spinor field. Having recovered the equivalence between these current spinor definitions, we undergo to the building blocks of the spinorial representation space, namely the Fierz aggregate and the bilinear identities, after what we define the space itself, allowing for the connection of its points to a physical spinor, regardless the chosen classical definition. 

\section{Lounesto's spinors classification, Pauli-Fierz-Kofink identities, and the Fierz aggregate}

Any spinor field
$\psi \in \sec \mathbf{P}_{\mathrm{Spin}_{1,3}^{e}}(M)\times_{\tau
}\mathbb{C}^{4}$ can be employed to construct its bilinear covariants  as section of bundle $\Omega(M)$,  reading \cite{LOU,Cra,bj}
\begin{subequations}\begin{eqnarray}
\label{cova}
\sigma&=&\bar\psi\psi\in\Omega^0(M),\\
 \omega&=&-\bar\psi\gamma_0\gamma_1\gamma_2\gamma_3\psi\in\Omega^4(M),\\
 \mathbf{J}&=&(\bar\psi\gamma_{\mu}\psi)\;\gamma^{\mu}\in\Omega^1(M), \\
\mathbf{K}&=&i(\bar\psi\mathrm{\gamma_0\gamma_1\gamma_2\gamma_3}\gamma_{\mu}\psi)\;\gamma^{\mu}\in\Omega^1(M),\\
 \mathbf{S}&=&\left(\bar\psi\,[\gamma_{\mu}, \gamma_{\nu}]\,\psi\right)\,\gamma^{\mu}\wedge\gamma^{\nu}\in\Omega^2(M)\,.
\end{eqnarray} \end{subequations}
 Equivalently, the components of the bilinear covariants are, respectively, 
denoted by
\begin{subequations}
\begin{eqnarray}
J_\mu &=& \bar\psi\gamma_{\mu}\psi,\\
K_\mu &=&i\bar\psi\gamma_{0}\gamma_{1}\gamma_{2}\gamma_{3}\gamma_{\mu}\psi,\\
S_{\mu\nu}&=&\bar\psi\,[\gamma_{\mu}, \gamma_{\nu}]\,\psi\,.
\end{eqnarray}
\end{subequations}

The bilinear covariants in the Dirac theory are interpreted respectively  as the mass term (or invariant length) in the Lagrangian of the electron ($\sigma$), the pseudo-scalar ($\omega$) relevant for parity-coupling, the current of probability density ($\mathbf{J}$), the chiral current density ($\mathbf{K}$), and the probability density of the (intrinsic) electromagnetic moment ($\mathbf{S}$) \cite{LOU,Cra}. 
A prominent requirement for the Lounesto's spinors classification is that the bilinear covariants satisfy quadratic algebraic relations, namely, the so-called Fierz-Pauli-Kofink (FPK) identities, which read
\begin{subequations}
\begin{eqnarray}
\label{Fierz}
\mathbf{J}^{2}&\equiv& J^\mu J_\mu =\omega^{2}+\sigma^{2},\\
\mathbf{K}^{2}&\equiv& K^\mu K_\mu =-\mathbf{J}^{2},\\
\mathbf{J}\cdot\mathbf{K}&\equiv& J^\mu K_\mu =0,\\
\mathbf{J}\wedge\mathbf{K}&
=&-(\omega+\sigma\gamma_{0}\gamma_{1}\gamma_{2}\gamma_{3})\mathbf{S}.  \label{fi}
\end{eqnarray}
\end{subequations}
When an arbitrary spinor $\xi$ satisfies $\gamma_0\xi^{\dagger}\gamma_0\psi\neq 0$, the original spinor $\psi\neq 0$ can be reconstructed, using the aggregate 
\begin{eqnarray}
\mathbf{Z}=\sigma + \mathbf{J} +i\mathbf{S}+i\mathbf{K}\gamma_0\gamma_1\gamma_2\gamma_3+\omega \gamma_0\gamma_1\gamma_2\gamma_3\, \label{Z}
\end{eqnarray}
by (a version) of the inversion theorem, $
\psi=\frac{1}{2\sqrt{\xi^{\dagger}\gamma_{0}\mathbf{Z}\xi}}\;e^{-i\theta}\mathbf{Z}\xi, \label{3}%
$  where
$\theta=i\log\left({2}(\xi^{\dagger}\gamma_{0}\psi{\xi^{\dagger}\gamma_{0}\mathbf{Z}\xi})^{-1/2}\right)$ is  \cite{Cra,Holland:1986wj,mosna}. 
   Moreover, when  $\sigma,\omega,\mathbf{J},\mathbf{S}%
,\mathbf{K}$ satisfy the Fierz identities, then the complex multivector
operator $\mathbf{Z}$ is named a {Fierz aggregate}, which can be self-adjoint, being called a
 boomerang in this case \cite{LOU}.
 The regular spinors are those whose at least one of the  bilinear covariants $\sigma$ and $\omega$ do not vanish. On the other hand, singular spinors present $\sigma=0=\omega$, and, in this case, the Fierz identities, given in Eq. \eqref{Fierz},  are in general replaced
by the most general conditions \cite{Cra}:
\begin{subequations}
\begin{align}
\!\!\!\!\!\frac14\mathbf{Z}^{2}    =\sigma \mathbf{Z},\qquad\qquad \frac14\mathbf{Z}\gamma_{\mu}\mathbf{Z}=J_{\mu}\mathbf{Z},\qquad\qquad \frac14\mathbf{Z}i[\gamma_{\mu},\gamma_{\nu
}]\mathbf{Z}=S_{\mu\nu}\mathbf{Z},\\ 
\frac14\mathbf{Z}i\gamma_0\gamma_1\gamma_2\gamma_3\gamma_{\mu}\mathbf{Z}    =K_{\mu}\mathbf{Z},\qquad\qquad -\frac14\mathbf{Z}\gamma_0\gamma_1\gamma_2\gamma_3\mathbf{Z}=\omega \mathbf{Z}.
\end{align}
\end{subequations}
The conditions are satisfied also by regular spinors. Insomuch, such relations are called by more general Fierz-Pauli-Kofink identities, however, written based on the Fierz aggregate.

Moreover, the inversion theorem (to be further regarded in the next section) is inspired by this spinor representation. More significantly here, the aggregate plays a central role within the Lounesto's classification since, in order to complete the classification itself, $\mathbf{Z}$ has to be promoted to a boomerang, satisfying $\mathbf{Z}^{2}=4\sigma\mathbf{Z}$. Obviously, for regular spinors the above condition is satisfied and $\mathbf{Z}$ is automatically a boomerang. However, for singular spinors case it is not so direct. Indeed, for singular spinors we must envisage the underlying geometric structure  to the multivector. From the geometric point of view, the following relations between the bilinear covariants must be fulfilled, in order to ensure that the aggregate is a boomerang: $\mathbf{J}$ field must be parallel to $\mathbf{K}$ and both are elements in the plane formed by the bivector field $\mathbf{S}$. Hence, using the Eq. (\ref{Z}) and taking into account singular spinors, it is straightforward to see that the aggregate can be recast as \cite{LOU}
\begin{eqnarray}
\mathbf{Z}=\mathbf{J}(1+i\mathbf{s}+ih\gamma_0\gamma_1\gamma_2\gamma_3), \label{ZB}
\end{eqnarray} where $\mathbf{s}$ is a space-like vector orthogonal to $\mathbf{J}$, and $h$ is a real scalar that is related to the spinor helicity. The multivector as expressed in Eq. (\ref{ZB}) is a boomerang \cite{Julio1}. The condition $\mathbf{Z}^{2}=4\sigma\mathbf{Z}$ yields $\mathbf{Z}^2=0$, for singular spinors, namely, the Fierz aggregate is nilpotent. However, for the FPK identities to hold,  the vector field $J$ must be light-like (isotropic) and the multivector $h\gamma_0\gamma_1\gamma_2\gamma_3+\mathbf{s}$ must be a pure imaginary \cite{LOU,Julio1}.

With these ingredients, it is possible to envisage six different classes of spinors, according to the following classification:
\begin{center}
\begin{tabular}[c]{||c|| c| c| c| c| c||}
\hline\hline
Class & $\sigma$ & $\omega$ & $\bf{K}$ & $\bf{S}$& $\bf{J}$\\
\hline\hline
1& $\neq 0$ & $\neq 0$ & $\neq 0$ & $\neq 0$& $\neq 0$\\
\hline
2& $\neq 0$ & $0$ & $\neq 0$ & $\neq 0$& $\neq 0$\\
\hline
3& 0 & $\neq 0$ & $\neq 0$ & $\neq 0$& $\neq 0$\\
\hline
4& 0 & 0 & $\neq 0$ & $\neq 0$& $\neq 0$\\
\hline
5& 0 & 0 & 0 & $\neq 0$& $\neq 0$\\
\hline
6& 0 & 0 & $\neq 0$ & 0& $\neq 0$\\
\hline\hline
\end{tabular}\medbreak
\footnotesize{Table 1:  Lounesto's spinor field classification.} 
\end{center}  The three first classes are composed by regular spinors, that comprise the standard textbook Dirac spinor. As stated in the literature, the representation spaces for the mentioned spinors are linked by the parity symmetry, however, quite recently regular spinors have been shown to be built without reference to this symmetry \cite{crd}. The elements of the fifth class are also called flag-pole spinors, represented by particular cases as Majorana and Elko  spinors,  whereas the sixth class comprises Weyl spinors. The fourth class, the flag-dipole, has had its first physical example recently discovered \cite{fabri}. For later reference we stress that $\mathbf{J}$ is always non null within this context. 
The Lounesto's classification has been explored in a comprehensive range of context, comprising field theory \cite{5307,daSilva:2012wp}, cosmology \cite{Fabbri:2010ws},  gravitation \cite{Ahluwalia} and formal aspects as well \cite{bonora,daRocha:2011yr,Julio1}. The general form of spinors in each one of the above classes was derived in Refs. \cite{Cavalcanti:2014wia,fabri}, and a classification that encodes gauge aspects was established in Ref. \cite{Fabbri:2016msm}.

\section{The Representation Space}

Bearing in mind that a given spinor can be written as a section of the bundle $\mathbf{P}_{\mathrm{Spin}_{1,3}^{e}}(M)\times _{\tau }\mathbb{C}^{4}$ we shall envisage the spinor space structure adopting a bottom-up, and somewhat pragmatic, approach by defining the regarded manifolds and spaces with respect to their points and elements. Notice that, as reinforced throughout Sect. II, the understanding of spinors as sections of the aforementioned bundle are not strictly necessary, although highly convenient as we shall see.   

In what follows let us denote by $\mathring{N}$ the five-dimensional manifold whose points are section  in sec $\Omega^a(M)$, with $a=0,\ldots,4$. The space $\mathring{N}$ is  isomorphic to the exterior bundle $\Omega(M)=\oplus_{a=0}^4\Omega(M)$.
 Let us denote by $P=(p^0,p^1,p^2,p^3,p^4)$ an arbitrary point of the manifold $\mathring{N}$, and the function $Z$, that establishes such a canonical isomorphism $\mathring{N}\overset{Z}{\simeq}\Omega(M)$. Obviously  $Z(P)\in \Omega(M)$.
 \medbreak
{\bf Definition 1:} \textit{$\mathring{\Sigma}$ is  the space whose elements are given by $Z\upeta$, where $\upeta \in \mathbb{C}^4$.}
\medbreak
Notice that as long as $Z$ is restricted to the bilinear covariants, namely,  we impose that it acts only upon points of $N$ satisfying the FPK identities, then the Fierz aggregate is straightforwardly obtained. Equivalently, however more generally, we proceed with the following direct construction:
\medbreak
{\bf Definition 2:} \textit{$N$ is a submanifold of $\mathring{N}$ whose points are such that $Z(P)$ obeys the FPK identities.}
\medbreak
When acting upon elements of $M$ it is convenient to write $Z(P)$ as
\begin{equation}
Z(P)=\sigma+{\bf J} + {\bf K} + {\bf S}+\omega,\nonumber
\end{equation} making explicit the multivector structure in terms of the bilinear covariants, just as to express $P=(\sigma, {\bf J}, {\bf K}, {\bf S},\omega)$.
\medbreak
{\bf Definition 3:} \textit{The representation space $\Sigma(N)$ is performed by elements given by $Z(N)\eta$, where $Z(N)$ stands for $Z(P)$ with $P\in N$ only. Therefore $Z\upeta\cong \Psi\in \Sigma(N)$ and the elements of $\Sigma$ are, thus, physical spinors.}
\medbreak
It is worth to emphasize that since the bilinears are invariant with respect to Lorentz transformation, the elements of $\Sigma(N)$ respect a relativistic dynamics. Clearly $\Sigma(N)\subset\mathring\Sigma$, i. e., the representation space is contained in the broader spinorial space. Therefore, the complement space $\mathring\Sigma\setminus\Sigma(N)$ comprises  points corresponding to spinors which \emph{do not} obey the FPK identities, the so-called anomalous spinors. 

The underlying idea to this construction regards the possibility to change from one physical spinor configuration to another one, by covering a given continuous path in the representation space. Differently of what happens to $\mathring{N}$, however, the submanifold $N$ must have a quite constraint topology inherited from the validity of the FPK identities.

Let us make this point clearer by considering merely regular spinors for a moment. In this case, it is possible to attain the appropriate subspace of $\Sigma(N)$ by defining the following canonical projector $\xi_{\rm reg}$:
\begin{eqnarray}
\xi_{\rm reg}: N&\rightarrow&\Omega(M)\nonumber\\(\sigma, {\bf J}, {\bf K}, {\bf S},\omega)&\mapsto& (\sigma, {\bf J}, 0, 0, \omega),\nonumber
\end{eqnarray} 
with image $\Omega^0(M)\oplus\Omega^1(M)\oplus\Omega^4(M)=\xi_{\rm reg}(N)$. 
Within the space $\xi_{\rm reg}(N) \subset \Sigma$, taking into account the identity ${\bf J}^2=\sigma^2+\omega^2$, that holds for regular spinors, it is always possible to associate a topological invariant for every regular state. Moreover, taking into account the usual mass dimension $3/2$ fermion, for which ${\bf J}$ represents the conserved current, the regarded topological invariant  must be related to the electric charge. Before proceeding, let us make two parenthetical remarks. First, it is straightforward to realize that $\xi_{\rm reg}$ may be naturally adapted for a lower dimension projection leading to elements as either $(\sigma, {\bf J}, 0, 0, 0)$ or $(0, {\bf J}, 0, 0, \omega)$. Second, when we refer to mass dimension of a given spinor, we mean the canonical mass dimension which shall be inherited by the quantum field from the dynamic respected by the expansion coefficients. Particular cases of the expansion coefficients are the objects treated here.

It is worth to emphasize that the space $\xi_{\rm reg}(N)$ has a rich underlying geometric structure. Indeed, it consists not merely a of submanifold, but furthermore it manifests a intriguing structure arising from the monopole construction of the Hopf
fibration $S^1\ldots S^3\rightarrow S^2$, where $S^1$ is homeomorphic to the
Lie gauge group U(1) of the electromagnetism \cite{jayme}. Using a similar construction,   the instanton is related to a principal bundle with structure
Lie group SU(2), homeomorphic to the 3-sphere $S^3$. The instanton
was described in Refs. \cite{jayme,jayme1} using the Hopf fibration $S^3\ldots
S^7\rightarrow S^4$, in the context of the Witten's monopole equations, by means of the bilinear covariants associated with regular spinor fields, under the Lounesto's spinor field classification \cite{Julio1}.

Let us make this point clear, working with a slightly different $\xi_{\rm reg}(N)$ space, after what the general case shall be regarded. Regular spinor fields in either class 1 or class 2  in Lounesto's classification can be thought of as satisfying $%
\sigma=1$ without loss of generality, defining the manifold $S^7$, when the Dirac spinor field is
classically described by an element of $\mathbb{C}^4\simeq\mathbb{H}^2$. Considering $\mathcal{C}\ell_4$ be the Clifford algebra of the 4-dimensional Euclidean vector space $\mathbb{R}^4$, i.e, the algebra generated by the set of vectors $\lbrace {\rm e}_{\mu}\rbrace$, subjected to the relations $
{\rm e}_{\mu}^2 = 1$, and ${\rm e}_{\mu}{\rm e}_{\nu} +{\rm e}_{\nu}{\rm e}_{\mu}=0$, with $ \mu = 0,1,2,3$.

The quaternionic representation can be constructed by using Eq. (\ref{quatrn}), by 
$
 {\rm e}_i \mapsto [e_i][e_0].
$ One defines the observables of $\psi \in \mathbb{C}^4$ in a 4-dimensional Euclidean space by the following expressions:
\begin{eqnarray}
\label{bili1}\!\!\!\!\!\!\sigma = \bar\psi\psi, \qquad \omega = \bar\psi{\rm e}_5\psi, \qquad J_{\mu} = \bar\psi{\rm e}_{\mu}\psi, \qquad K_{\mu} = i\bar\psi{\rm e}_5{\rm e}_{\mu}\psi, \qquad S_{\mu\nu} = \bar\psi[{\rm e}_{\mu},{\rm e}_{\nu}]\psi,
\end{eqnarray}
where ${\rm e}_5 = {\rm e}_0{\rm e}_1{\rm e}_2{\rm e}_3$. 
Note that the only different relations due to the Minkowski space case is the expression for $\omega$ (see Eq.\eqref{cova}). In fact, in the $\mathcal{C}\ell_{1,3}$ algebra one has ${\rm e}_5^2 = -1$,  whereas considering $\mathcal{C}\ell_{4}$ yields  ${\rm e}_5^2=1$. 
In this way, the Fierz identities must be modified \cite{jayme,mosna}, yielding 
\begin{eqnarray}\label{if}
\boldsymbol{{\rm J}}^2 = \sigma^2 - \omega^2, \qquad \boldsymbol{{\rm J}}^2=\boldsymbol{{\rm K}}^2, \qquad \boldsymbol{{\rm J}}\wedge\boldsymbol{{\rm K}} = (\sigma - {\rm e}_5\omega)\boldsymbol{{\rm S}}, \qquad J^\mu K_\mu = \boldsymbol{{\rm J}}\cdot\boldsymbol{{\rm K}} = 0.
\end{eqnarray}
In this context, the first Fierz identity \eqref{if} provides the expression $\mathbf{J}^2 + \omega^2 =1$ defining the 4-sphere $S^4$,  with 
coordinates $(J_\mu,\omega)$.

Now, taking into account Eqs. (\ref{hh})  and  Eq.(\ref{cova}) yields \cite{jayme,jayme1}
\begin{eqnarray}
\sigma &=& q_1\cdot q_1 \!+\! q_2\cdot q_2,\qquad \omega = 2\,\Re(q_1^\ast
q_2),\qquad J_0 = q_1\cdot q_1 \!-\! q_2\cdot q_2, \qquad 
   J_i = 2\,\epsilon_i^{\;jk}\Re(q_1^\ast\,{\rm e}_j{\rm e}_k\, q_2), \qquad 
\end{eqnarray}
\noindent \text{for $i,j,k=1,2,3$} and $\epsilon_i^{\;jk}$ is the Levi-Civita symbol. Hence, the representation in Eq. (\ref{bili1}) reads
\cite{jayme,jayme1}
\begin{eqnarray}
\sigma &=& |\,\psi_1|\,^2 + |\,\psi_2\,|^2 + |\,\psi_3\,|^2 + |\,\psi_4\,|^2\,,\qquad \omega = 2 \Re(\psi_1\psi_3^\ast + \psi_2\psi_4^\ast), \\
J_0 &=& |\,\psi_1\,|^2 + |\,\psi_2\,|^2 - |\,\psi_3\,|^2 - |\,\psi_4\,|^2  \notag, \qquad
J_1 = 2 \Im(\psi_1\psi_4^\ast +\psi_2\psi_3^\ast)\,,\\
J_2 &=& 2 \Re(\psi_2\psi_3^\ast - \psi_1\psi_4^\ast),
\notag \qquad J_3 = 2 \Im(\psi_3\psi_1^\ast +\psi_2\psi_4^\ast)\,.  \label{s44}
\end{eqnarray}%
\noindent
This important geometric feature reveals that
the underlying geometry induced by spinors classes can further point to more structures. Indeed, the following proposition regards the topological invariants associated to regular spinors.
\medbreak

{\bf Proposition 1}: \textit{Let $\xi_{\rm reg}(N)$ be the space consisting of regular spinors. Then $$\pi_{[\dim(\xi_{\rm reg}(N))-1]}(\xi_{\rm reg}(N))=n\in\mathbb{Z},$$ where $\pi_i(R)$ stands for the $i^{\rm th}$-homotopy group of a given $R$ space. In the case of a mass dimension $3/2$ fermion described by the regular spinor, the topological invariant is associated to the electric charge.}
\medbreak
{\bf Proof}: The first assertion directly follows  from the above discussion. In fact, a regular spinor is an element of $\xi_{\rm reg}(N)$ and therefore there exists just two possibilities:
\begin{itemize}
\item $\dim\xi_{\rm reg}(N)=3$, and the spinor belongs to class 1 of Lounesto's classification. In this case the point $(0,0,0)\in\xi_{\rm reg}(N)$ can not be attained, implying that $\pi_2(\xi_{\rm reg}(N))=n\in\mathbb{Z}$;

\item $\dim\xi_{\rm reg}(N)=2$, and thus the spinor belongs to either class 2 or class 3 of Lounesto's classification. This is the case when $\pi_1(\xi_{\rm reg}(N))=n\in\mathbb{Z}$.
\end{itemize}
In general, ${\bf J}$ is not necessary related to the conserved current associated to a given fermion. Nevertheless, when it does -- what is the case for mass dimension $3/2$ fermions -- then the conserved charge is the electric charge itself. Since a vanishing charge is forbidden in these situations, one has the physical counterpart of the above topological constraint. Indeed the $(0,0,0)$ point for  regular spinors can never be reached.   \hspace{1cm}$\Box$
\medbreak
As it is readily verified, in the case of the three dimensional $\xi_{\rm reg}(N)$ space the conserved current may also be seen as the generator of cohomology, for $H^1(\xi_{\rm reg}(N))\simeq H^1(\mathbb{R}^2\setminus\{0\})$ and the usual closed form
\begin{eqnarray}
\frac{\sigma d\omega-\omega d\sigma}{\sigma^2+\omega^2},
\end{eqnarray} is not everywhere exact. The usefulness of the representation space construction is now evident. By treating physical spinor (states) as points of a given space, constrained by the algebraic bilinear relations, it is possible to work in the interplay of topology, multivector algebra, and physics. As a matter of fact, the very existence of a relativistic spinor is related to a topological invariant in the representation space. 

Nevertheless, one shall not be so optimistic just by looking at the example just studied, since we were dealing only with a projection. The $\Sigma$ space, where not only regular spinors are taken into account, is certainly very difficult to be analysed. There are, however, some interesting points that we shall report on the study of $\Sigma$ in its general form. In fact, Lounesto's classification provides six classes of spinors, wherein a continuous path in the representation space allows access to different configuration states. Let us make this idea more clear and precise.

All spinors in an arbitrary class  are connected by a simple rescaling. From the point of view of elements in $\Sigma(N)$, two different elements $\Psi'$ and $\Psi$ are connected by an usual transformation along the same class by $\Psi'=S\Psi$. In this context, it is possible to assert, in a manner akin to Wigner \cite{Wigner:1939cj}, the following proposition.
\medbreak
{\bf Proposition 2}: \textit{Let $D_{\lambda}$ be an 1-parameter infinitesimal operator acting on the spinor space of a given class according Lounesto's classification. Suppose that it is an homomorphism, $D_\lambda D_{\lambda'}=D_{\lambda+\lambda'}$, with $\lambda \in \mathbb{R}$. If there exists a physical state on which the application of $D_\lambda$ is well defined, then there exists a dense set of such states in the respective class, with respect to the Lounesto's classification.}
\medbreak
{\bf Proof}: If the application of $D_\lambda$ is well defined for a given state, there exists the limit $ \lim_{\lambda\rightarrow 0} \lambda^{-1}(D_{\lambda}-1)\Psi,$ what implies that $\lim_{\lambda\rightarrow 0} \lambda^{-1}(D_{\lambda}S^{-1}S-S^{-1}S)\Psi$. Since  the rescaling commutes with $D_\lambda$, it follows that  there exists the limit $ \lim_{\lambda\rightarrow 0} \lambda^{-1}(D_{\lambda}-1)S\Psi.$ Hence, within an arbitrary but fixed class in Lounesto's classification, it is possible to operate with infinitesimal operators in a rather usual way. Moreover, in view of the above result, physical spinors are indeed points of $\Sigma(N)$. \hspace{1cm}$\Box$

It is important to remark that an arbitrary class in Lounesto's classification is invariant under $S$. It is furthermore possible, however, to connect two different classes by algebraic transformation. More specifically, it was shown in Ref. \cite{JMP2007} that there exists a subset of spinors in class 1, 2, and 3 which can be mapped into a subclass of class 5 spinors. Let us denote this transformation,  between different classes, by $S_{C}$. It turns out that $\det S_C\neq 0$ \cite{JMP2007}. Hence, the alluded algebraic bridge, in a manner of speaking, is also dense. In fact, as far as we restrict ourselves to the subset of states which can be mapped, the proof of Proposition 2 holds, in this switching class case. 

A given algebraic bridge, however, is not always necessarily well behaved. In the sequel we give a (counter-) example, presenting a mapping between a subset of spinors in class 1, 2, and 3, and a subset in class 4, which is neither Hermitian nor invertible. As one shall see, this example is quite severe in the constraints it imposes,  and it is not discarded the possibility of a more manageable mapping. It is worth to  stress that the mapping, from regular spinors to class 4 spinors, is chosen as a particular case. Class 4 spinors are understood as the most unvoiced class in Lounesto's classification, having just a rare single example in the literature \cite{fabri} as a physical solution of the Dirac equation in a Riemann-Cartan Bianchi-I, $f(R)$, background. Lounesto  describes such class as the only one that, at that time, had not corresponded to any type of spinor already found in Nature \cite{LOU}.  Except for such solution, neither other types of flag-dipole spinors nor their respective dynamics as well have been found, yet. The algebraic mapping between regular and class 4 spinors can be parenthetically seen, then, as an attempt to put forward a bottom-up approach, embracing flag-dipoles spinors into the standard setup of high energy physics.  

We start by introducing a matrix $\textit{M}={\small\begin{pmatrix}
\mathbb{M}_{\mathrm{11}} & \mathbb{M}_{\mathrm{12}}\\
\mathbb{M}_{\mathrm{21}} & \mathbb{M}_{\mathrm{22}}
\end{pmatrix}} \in \mathcal{M}(4,\mathbb{C})$, defining the mapping:
\begin{eqnarray}
M: \mathcal{D} &\rightarrow& \mathcal{T}_{\mathrm{4}}\nonumber\\
\Phi_{D}&\mapsto& \Psi_{4}=\textit{M}\Phi_{D},\label{MAPA}
\end{eqnarray} where $\mathcal{D}$ and $\mathcal{T}_{\mathrm{4}}$ stands for the sets comprising regular and flag-dipole spinors, respectively. The formalism is clearly 
representation-independent, however the Weyl representation 
$\footnotesize{\gamma_{0}={}
\begin{pmatrix}
\mathbb{O} && \mathbb{I} \\
\mathbb{I} && \mathbb{O}
\end{pmatrix}}$, $ 
\footnotesize{
\gamma_{k}=
\begin{pmatrix}
\mathbb{O} && \sigma_{k} \\
-\sigma_{k} && \mathbb{O}
\end{pmatrix}}$ shall be used, being $\sigma_k$ the usual Pauli matrices, to fix the notation hereon. 

According to the Lounesto's spinor classification, using the mapping defined in (\ref{MAPA}), class-4 spinors satisfy
\begin{eqnarray}
\label{16}
\sigma=\Phi_{D}^{\dag}\textit{M}^{\dag}\gamma_{0}\textit{M}\Phi_{D}=0, \;\;\;\;\;\; \qquad \omega=-\Phi_{D}^{\dag}\textit{M}^{\dag}\gamma_{123}\textit{M}\Phi_{D}=0.
\end{eqnarray} Let us investigate here the constraints exclusively on the $\textit{M}$ matrix. The conditions (\ref{16}) imply
\begin{eqnarray}
\label{12}
\textit{M}^{\dag}\gamma_{0}\textit{M}=0,\qquad\textit{M}^{\dag}\gamma_{123}\textit{M}=0,
\end{eqnarray} which, in the Weyl representation,  yield
\begin{eqnarray}
\label{7}
\pm\mathbb{M}^{\mathrm{\dag}}_{\mathrm{11}}\mathbb{M}_{\mathrm{21}}+\mathbb{M}^{\mathrm{\dag}}_{\mathrm{21}}\mathbb{M}_{\mathrm{11}}=\mathbb{O}=
\pm\mathbb{M}^{\mathrm{\dag}}_{\mathrm{11}}\mathbb{M}_{\mathrm{22}}+\mathbb{M}^{\mathrm{\dag}}_{\mathrm{21}}\mathbb{M}_{\mathrm{12}},\\
\label{9}
\pm\mathbb{M}^{\mathrm{\dag}}_{\mathrm{12}}\mathbb{M}_{\mathrm{21}}+\mathbb{M}^{\mathrm{\dag}}_{\mathrm{22}}\mathbb{M}_{\mathrm{11}}=\mathbb{O}=
\pm\mathbb{M}^{\mathrm{\dag}}_{\mathrm{12}}\mathbb{M}_{\mathrm{22}}+\mathbb{M}^{\mathrm{\dag}}_{\mathrm{22}}\mathbb{M}_{\mathrm{12}}\,.
\end{eqnarray}
The system (\ref{7})-(\ref{9}) is satisfied when
\begin{eqnarray}
\label{17}
\mathbb{M}^{\dag}_{\mathrm{11}}\mathbb{M}_{\mathrm{21}}=
\mathbb{M}^{\dag}_{\mathrm{11}}\mathbb{M}_{\mathrm{22}}=
\mathbb{M}^{\dag}_{\mathrm{12}}\mathbb{M}_{\mathrm{21}}=
\mathbb{M}^{\dag}_{\mathrm{12}}\mathbb{M}_{\mathrm{22}}=\mathbb{O}.
\end{eqnarray} Therefore, by considering a general representation, 
\begin{center}
$\mathbb{M}$$_{11}={}
\begin{pmatrix}
m_{11} && m_{12}\\
m_{21} &&m_{22}
\end{pmatrix}
{},\;\;$
$\mathbb{M}$$_{12}={}
\begin{pmatrix}
m_{13} && m_{14}\\
m_{23} &&m_{24}
\end{pmatrix}
{},\;\;\mathbb{M}$$_{21}={}
\begin{pmatrix}
m_{31} && m_{32}\\
m_{41} &&m_{42}
\end{pmatrix}
{},$\;\;
$\mathbb{M}$$_{22}={}
\begin{pmatrix}
m_{33} && m_{34}\\
m_{43} &&m_{44}
\end{pmatrix}
{},$
\end{center} 
it is possible to rewrite 
\begin{equation}
\label{13}
\textit{M}={}
\begin{pmatrix}
m_{11} & m_{12} & m_{13} & m_{14}\\
\frac{m_{11}m_{22}}{m_{12}} & m_{22} & \frac{m_{13}m_{22}}{m_{12}} &\frac{m_{14}m_{22}}{m_{12}}\\
\frac{-m_{22}^{*}m_{41}}{m_{12}^{*}} & \frac{-m_{22}^{*}m_{42}}{m_{12}^{*}} & \frac{-m_{22}^{*}m_{43}}{m_{12}^{*}} &\frac{-m_{22}^{*}m_{44}}{m_{12}^{*}}\\
m_{41} & m_{42} & m_{43} &m_{44}
\end{pmatrix}
{}.
\end{equation}

The $\textit{M}$ matrix is responsible to perform the mapping between regular and class-4 spinors\footnote{A simple, but tedious calculation show that the other bilinears behave in such a way that the mapping (\ref{13}) works well, ensuring a final class-4 spinor.}. There are, however, important restrictions on such a mapping which must be highlighted. 

Firstly, the mapping performed cannot occur from class-4 to regular spinors. In fact, supposing the existence of ${\mathring{M}}$ such that $\mathring{{M}}\Psi_{4}=\Phi_{D}$, then ${M}\mathring{{M}}\Psi_{4}=\textit{M}\Psi_4$, leading to $\mathring{{M}}={M}^{-1}$. Nevertheless, as it can be explicitly calculated from (\ref{13}), $\det{{M}}=0$ and there is no such a $\mathring{{M}}$ matrix. Secondly, the mapping (\ref{13}) cannot be Hermitian. Indeed, the requirement $\textit{M}=\textit{M}^\dagger$ yields 
\begin{eqnarray}
m_{11}&=&m_{11}^{*}, \;\;\;\;\;\;\; m_{12}=m_{12}^{*},\;\;\;\;\;\;\;m_{22}=m_{22}^{*},\;\;\;\;\;\;\;\ m_{14}=m_{41}^{*}, \\
m_{13}&=&\frac{-m_{22}m_{14}}{m_{12}}=-m_{42}^{*},\;\;\;\;\;\;\; m_{44}=\frac{-m_{12}m_{43}}{m_{22}}.
\end{eqnarray} Nevertheless, the other bilinear invariants do not behave as the ones for class-4 spinors. In fact, it can be verified that ${\bf K}=0=\bf{S}$ for this case, rendering a spinor different from a class-4 one and, then, the hermiticity is forbidden. 

It is significant to stress that the above mapping was performed using class-1 regular spinors, as it is clear from (\ref{MAPA}) and (\ref{16}). Nevertheless, as far as we implement the additional constraints coming from class-2 and 3 Dirac-like spinors, the final form of the bilinear invariants are slightly modified, but the net result is the same. While property one is sound in prospecting possible information about the representation space (in the context of Proposition 2), the study of the hermicity property  may be relevant in a quantum mechanical context.   

The counter-example just studied indicates an elaborated representation space, whose non-triviality deserves further exploration. It must be once again emphasized, however, that the constraints (\ref{12}) are too restrictive, since it extends the kernel of the transformation to the whole $(\sigma, \omega)$-plane. 

 We would like to finalize this work by pointing out three new classes of spinors which also reside in the spinor representation space. These spinors were obtained in the operatorial and algebraic form in Ref. \cite{Villalobos:2015xca}, having, by construction, $\mathbf{J}=0$. Therefore, their dynamics cannot be described by the Dirac operator. The nontrivial topology of the representation space, as already remarked, is inherited from the constraints imposed by the FPK identities. For the sector of $\Sigma(N)$ comprised by regular spinors, $\mathbf{J}$ is the generator of cohomology and cannot vanish. The sector of $\Sigma(N)$ encompassing singular spinors, nevertheless, may also accommodate the spinors of \cite{Villalobos:2015xca}. Notice that a vanishing $\mathbf{J}$ does not lead to a contradiction, and the FPK identities still hold in this case. Hence,  these spinors are also physical in the sense previously discussed. The spinors founded in \cite{Villalobos:2015xca}, assuming that ${\bf
 J}=0$,  may be called pole (only $\mathbf{K}\neq 0$), flag (only $\mathbf{S}\neq 0$), or flag-pole\footnote{It is worth to mention 
 that these flag-poles are essentially different 
 of the standard flag-poles characterized by the flag $\mathbf{S}\neq 0$) and the pole ($\mathbf{J}\neq 0$), since in this case $(\mathbf{K}= 0)$ \cite{LOU}.}. They live in a special subspace of $\Sigma(N)$ whose topology also deserve further attention. 

\section{Concluding Remarks}

The formalization of a spinor representation space, whose points can be faced as physical spinors, has been constructed. These spinors have been shown to behave into dense paths of the representations space which, in view of the FPK identities, perform highly topologically constrained subsets. Some of these subsets have topological properties intrinsically connected to physical relevant quantities. The representation space show itself as an adequate tool to explore dynamics and interactions usually by means of using infinitesimal operators.

It should be emphasized that along this work we took advantage from dealing with spinors as elements of $\mathcal{C}\ell_{1,3}\frac{1}{2}(1+{e}_0)$ in Sec. II. Similar constraints in the representation space, coming from the FPK identities in the $\mathcal{C}l_4$ isomorphic case, are expected. However, our main interest here is the study of the representation space taking into account the Clifford algebra constructed upon the Minkowski space.

In showing that type-4 spinors can not be led into regular ones, we asserted about  the mapping connecting different physical spinors -- spinors of different sectors do $\Sigma(N)$, belonging to different classes. However, no reference has been made about quantum mechanics. It is time to elaborate this a little further. The very possibility of crossing over different classes, by means of a well defined algebraic transformation connecting different sectors of $\Sigma(N)$ could, in principle, be related to some type of swapping spinor class due to a specific physical process. In fact, bearing in mind the existence of a dense set in between different classes in the light of proposition two, this switching could be performed by a specific (unknown) scattering matrix modeling the physical process.  Apart from unitarity concerns\footnote{Typically, the $S_C$ matrix have enough symmetry to be recast into a specific form allowing for an unitary scattering process}, it is difficult to envisage how this attemptive process can duplicate the helicity states in going from regular spinors to type-5 spinors, these last spinors with known dual property helicity. Perhaps, and here we are entering in the fancy ground of speculation, a comprehensive transformation performed in the quantum operator as a whole may seed a precise light into the formal aspect of this possible swapping. It turns out, however, that the physical process would still be lacking. In the opposite way of this expectation, we delivered a mapping which is neither invertible nor Hermitian, evincing the high degree of topological constraint presented on the representation space. Further investigation on the algebraic/topological relationship concerning singular spinors are under current investigation. 

\section*{Acknowledgments}
JMHS thanks to CNPq (304629/2015-4; 445385/2014-6) for partial financial support. CHCV thanks to CAPES (PEC-PG) for financial support. RJBR thanks to CAPES for financial support, and RdR~is grateful to CNPq (Grant No. 303293/2015-2),
and to FAPESP (Grant No.~2015/10270-0), for partial financial support.

\end{document}